\documentclass{article}
\usepackage{subfigure}   
\makeindex
\usepackage{amsmath}
\usepackage{graphicx}
\usepackage{caption}
\usepackage[normalem]{ulem}
\usepackage[pdfpagelabels=false]{hyperref}  
\usepackage{color}
\usepackage{authblk}
\newcommand{\fm}{{\rm\,fm}}
\newcommand{\MeV}{{\rm\,MeV}}
\newcommand{\fmd}{{\rm\,fm^{-3}}}

\newcommand{\fmk}{{\rm\,fm^{-1}}}
\newcommand{\I}{\mathrm{i}}
\renewcommand{\vec}[1]{\mbox{\boldmath $#1$}}
\newcommand{\vecs}[1]{\mbox{\boldmath \scriptsize$#1$}}
\begin{document}
\title{Clustering and pasta phases in nuclear density functional theory}
\author{Bastian Schuetrumpf, Chunli Zhang}
\affil{FRIB Laboratory, Michigan State University, East
  Lansing, Michigan 48824, USA}
\author{Witold Nazarewicz}
\affil{Department of Physics and
  Astronomy and FRIB Laboratory, Michigan State University, East
  Lansing, Michigan 48824, USA\\
  Faculty of Physics, University of Warsaw, 02-093 Warsaw, Poland}
  \maketitle
\begin{abstract}
Nuclear density functional theory (DFT) is the tool of choice in describing properties of complex nuclei and intricate phases of bulk nucleonic matter. It is a microscopic approach based on an energy density functional representing the nuclear interaction. An attractive feature of nuclear DFT is that it can be  applied to both finite nuclei and  pasta phases appearing in the inner crust of neutron stars. While nuclear pasta clusters in a neutron star can be easily characterized through their density distributions, the level of clustering of nucleons in a nucleus can often be difficult to assess. To this end, we use the concept of nucleon localization. We demonstrate that the localization measure provides us with fingerprints of clusters in light and heavy nuclei, including fissioning systems.  Furthermore we investigate the rod-like  pasta phase using twist-averaged boundary conditions, which enable calculations in finite volumes accessible by state of the art DFT solvers. 
\end{abstract}

\section{Introduction}

Nuclear clustering is a ubiquitous phenomenon \cite{Beck1,Beck2,Beck3,Oertzen06,Cluster12}, but its comprehensive understanding still eludes us. On the one hand, the  very occurrence of cluster states  at low excitation energies around  cluster-decay thresholds \cite{Ikeda68} is a consequence of an openness of the nuclear many-body system \cite{Okolowicz12}. On the other hand, the properties of cluster states, once they are formed,  can be well described within a  mean-field picture \cite{Leander,Flocard84,Freer95,Ichikawa11,Ebran12,Ebran2014}, and their characteristics can be traced back to the symmetries and geometry of the nuclear mean-field \cite{Hecht77,Nazarewicz1992}. 
The quantitative theoretical description of clustering requires the use of advanced many-body, open-system  framework employing realistic interactions, and there has been a significant progress in this area \cite{Cluster12,(Elh15)}. If one is aiming at a global characterization of cluster states throughout the nuclear landscape and for neutron stars, a good starting point is  nuclear density functional theory (DFT) \cite{Bender03} based on an energy density functional (EDF) representing the effective in-medium nuclear interaction.

The degree of clustering  in nuclei is difficult to assess quantitatively in DFT
as the single particle wave functions are generally spread throughout the nucleus and the resulting  nucleonic distributions are rather crude  indicators of cluster structures. Therefore, in this study, we utilize a different measure, called spatial localization, which is a better signature of   clustering and cluster shell structure. The localization, originally introduced in the context of many-electron systems \cite{Becke,Savin97,Scemama,Burnus05}, has recently been applied to nuclear structure elucidations \cite{Reinhard2011}. 

The range of clustering phenomena in nuclear physics extends well beyond finite nuclei. In compact stellar objects such as neutron stars or supernova,  nuclear matter is present on the macroscopic scale. Especially remarkable are the phases of bulk nucleonic matter just below the nuclear saturation density. Here nuclear matter can lump into clusters, which have shapes of rods, slabs, and other complex, often very intricate, geometries. Those structures are usually referred to as nuclear pasta \cite{Ravenhall,Hashimoto}. 

Nuclear pasta phases have been studied in semi-classical models \cite{Dorso2012, Williams1985, Oka13a, Pais15, Schneider13, Schneider14, Maruyama, Sonoda2008, Watanabe09}, as well as in quantum mechanical models, including DFT with and without time-dependence \cite{Bonche, Mag02, Goegelein, NewtonStone, Pais12, Schuetrumpf2013a, Schuetrumpf2014, Schuetrumpf2015}. In this work, we study nuclear rods with twist-averaged boundary conditions (TABC) \cite{Gros1992,Gros1996,Lin}, which provide the most general solutions for states in a periodic potential. TABC minimize finite-volume effects drastically and enable precise calculations within the chosen model. TABC have been utilized in the context of the crust of neutron stars within DFT and  Quantum Monte Carlo approaches  \cite{Carter,Chamel2007,Schuetrumpf2015b,Gulminelli}.

This paper is organized as follows. We outline the nuclear DFT framework used in Sec.~\ref{sec:DFT}.  Section~\ref{sec:loc} defines the spatial localization of nucleons. In Sec.~\ref{sec:TABC} we describe the DFT+TABC approach in the context of infinite systems. In Sec.~\ref{sec:finitenuc} we apply the localization measure to finite nuclei, both light and heavy, and compare the DFT results with those obtained within the harmonic oscillator model. Section \ref{sec:nucrods} shows  the DFT+TABC predictions for the rod phase in the neutron star crust. We summarize our results and present the
outlook for the future in Sec.~\ref{Conclusions}.

\section{Nuclear Density Functional Theory\label{sec:DFT}}

Nuclear DFT models, including self-consistent mean-field models based on the Hartree-Fock (HF) or Hartree-Fock-Bogoliubov (HFB) approximation, are designed to describe nuclei all over the nuclear chart. The HFB equations are derived from the variational principle and the resulting nuclear wave function is represented by a product state.  In DFT, the interaction between the nucleons is approximated by an EDF, which can be written  as an expansion in local one-body densities and currents, such as, e.g., the particle density, kinetic density, and spin-orbit density. The EDF parameters have to be optimized to selected experimental data. In this work we use Skyrme type \cite{Bender03} EDFs. For finite nuclei,  we employ the optimized  parametrizations SkM* \cite{SkMS} ($^{264}$Fm and $^{132}$Sn) and UNEDF1-HFB \cite{UNEDF1HFB} (light nuclei, no pairing). For the nuclear rod calculations we compare the SLy6 \cite{Chabanat} and TOV-min \cite{Erler2013} parametrizations, both of which were designed to perform well for neutron-rich matter. 

For finite nuclei, we utilize the DFT solvers HFBTHO \cite{Stoitsov2013} and HFODD \cite{Schunck2012}. Both codes expand the self-consistent wave functions in a harmonic oscillator basis. While with HFODD all self-consistent mean-field symmetries can be broken, HFBTHO assumes axial geometry. For infinite systems, the harmonic oscillator representation of wave functions cannot be used. Here we use the symmetry-unconstrained DFT solver Sky3d \cite{Mar15a} based on  the fast Fourier transform method, in which the wave functions are represented on an equidistant Cartesian 3D grid.

\section{Spatial Localization\label{sec:loc}}

The localization measure has been originally introduced to characterize chemical bonds~\cite{Becke,Savin97,Scemama,Burnus05} and has also been demonstrated to be useful for characterizing clusters in light nuclei  \cite{Reinhard2011}. The localization can be defined through a probability  $ R_{q\sigma}(\vec{r},\delta)$ of finding a pair of nucleons at point $\vec{r}$ having  the same spin $\sigma$  and isospin $q=n,p$ within a small radius $\delta$. As discussed in \cite{Becke}, this probability can be written as:
\begin{equation}
R_{q\sigma}(\vec{r},\delta)\approx{1\over 3}\left(\tau_{q\sigma}-{1\over 4}\frac{[\vec{\nabla}\rho_{q\sigma}]^2}{\rho_{q\sigma}}-\frac{\vec{j}^2_{q\sigma}}{\rho_{q\sigma}}\right)\delta^2+\mathcal{O}(\delta^3)\label{eq:prob}
\end{equation} 
where the particle density $\rho_{q\sigma}$,  kinetic density $\tau_{q\sigma}$,  density gradient $\vec{\nabla}\rho_{q\sigma}$, and the current density $\vec{j}_{q\sigma}$ are defined as:
\begin{subequations}
\begin{eqnarray}
\rho_{q\sigma}(\vec{r})&=&\sum_{\alpha\in q}v^2_{\alpha}|\psi_\alpha(\vec{r}\sigma)|^2\\
\tau_{q\sigma}(\vec{r})&=&\sum_{\alpha\in q}v^2_{\alpha}|\vec{\nabla}\psi_\alpha(\vec{r}\sigma)|^2\\
\vec{\nabla}\rho_{q\sigma}(\vec{r})&=&2\sum_{\alpha\in q}v^2_{\alpha}\mathrm{Re}[\psi^*_\alpha(\vec{r}\sigma)\vec{\nabla}\psi_\alpha(\vec{r}\sigma)]\\
\vec{j}_{q\sigma}(\vec{r})&=&\sum_{\alpha\in q}v^2_{\alpha}\mathrm{Im}[\psi^*_\alpha(\vec{r}\sigma)\vec{\nabla}\psi_\alpha(\vec{r}\sigma)],
\end{eqnarray}
\end{subequations}
with $v^2_{\alpha}$ being the occupations of canonical orbits $\psi_\alpha(\vec{r}\sigma)$.
Henceforth, we take the expression in parentheses  of Eq.~(\ref{eq:prob}) as a localization measure. This expression is neither dimensionless nor normalized. A natural choice for normalization is the Thomas-Fermi kinetic density $\tau^\mathrm{TF}_{q\sigma}={3\over 5}(6\pi^2)^{2/3}\rho_{q\sigma}^{5/3}$. With this normalization, one defines a reversed and normalized localization measure,
\begin{equation}
\mathcal{C}_{q\sigma}(\vec{r})=\left[1+\left(\frac{\tau_{q\sigma}\rho_{q\sigma}-{1\over 4}|\vec{\nabla}\rho_{q\sigma}|^2-\vec{j}^2_{q\sigma}}{\rho_{q\sigma}\tau^\mathrm{TF}_{q\sigma}}\right)^2\right]^{-1},
\label{eq:loc}
\end{equation}
known for electronic systems as the electron localization function (ELF). In the following, we shall refer to $\mathcal{C}_{q\sigma}(\vec{r})$ as the nucleon localization function (NLF). A large value of NLF indicates that the probability of finding two nucleons with the same spin and isospin at the same location  is very low. Thus the nucleon's localization is large at that point. In particular the alpha particle has perfect localization for all combinations of spin and isospin  \cite{Reinhard2011}. The value of $\mathcal{C}=1/2$ is characteristic of a nearly homogeneous Fermi gas as found in nuclear matter. 
 
The above  definition of the NLF works  well in regions with non-zero nucleonic density. When the local densities become very small in the regions outside the range of the nuclear mean field, numerical artifacts become dominant, because the numerator and denominator in Eq.~(\ref{eq:loc}) are both close to zero (see, e.g.,  Fig.~1 of Ref.~\cite{Reinhard2011}). Consequently, for finite nuclei, we multiply the NLF by a normalized particle density $\mathcal{C}(\vec{r})\rightarrow\mathcal{C}(\vec{r})[\rho_{q\sigma}(\vec{r})/\mathrm{max}(\rho_{q\sigma}(\vec{r}))]$. Such a procedure is not necessary for the nuclear pasta case, because particle densities do not vanish in the regions  between the nuclear clusters in pasta phases.

\section{Twist-Averaged Boundary Conditions\label{sec:TABC}}
The code Sky3d is based on Fourier transforms for derivatives. Therefore the native boundary conditions are periodic boundary conditions (PBC). While PBC are well suited for densities and fields for most nuclear pasta calculations, like the nuclear rods discussed in Sec. \ref{sec:nucrods}, they represent a constraint on the HF single-particle states. According to the Floquet-Bloch theorem, the most general wave function in a periodic potential can be written as:
\begin{equation}
\psi_{\alpha\vecs{q}}(\vec{r})=u_{\alpha\vecs{q}}(\vec{r})e^{\I\vecs{q}\vecs{r}},
\end{equation}
where  $\psi_{\alpha\vecs{q}}(\vec{r})$ is the single-particle wave function labeled by quantum numbers $\alpha$, $\vec{q}$ is the wave vector that determines the boundary condition, and $ u_{\alpha\vecs{q}}(\vec{r}) $ is a periodic function of $\vec{r}$. These wave functions obey the boundary conditions 
\begin{equation}\label{bbc}
\psi_{\alpha\vecs{\theta}}(\vec{r}+\vec{T}_i)=e^{\I\theta_i}\psi_{\alpha\vecs{\theta}}(\vec{r}),
\end{equation}
where $\vec{\theta}$ is the set of Bloch angles (or twists) and $\vec{T}_i$ $(i=1,2,3)$ stands for the three lattice vectors.

In the DFT+TABC approach, the expectation value of an observable $\hat{O}$ is obtained by
 averaging over the twists:
\begin{equation}
\langle \hat{O}\rangle= \int  \frac{d^3\vec{\theta}}{\pi^3}\,\langle \Psi_{\vecs{\theta}}| \hat{O}| \Psi_{\vecs{\theta}}\rangle,
\label{eq:average}
\end{equation}
where $\Psi_{\vecs{\theta}}$ is the HF/HFB  state. The twists $ \theta_i $ change  between zero (PBC) and $\pi$ (anti-PBC), as the time-reversal symmetry is assumed \cite{Lin}. In this work, we use the DFT+TABC implementation described in Ref.~\cite{Schuetrumpf2015b}.

\section{Finite nuclei\label{sec:finitenuc}}

\subsection{Harmonic oscillator model}

Because of the large binding energy of $^4$He, $\alpha$-cluster states are prominent in alpha-conjugate nuclei. For those light systems, the deformed harmonic oscillator (HO)  model can serve as a rough guidance \cite{Leander,Nazarewicz1992,Freer95,Rae88}. To this end, we first study the NLF using the wave functions of the axial HO  with the deformation parameter $\eta=\omega_\bot/\omega_z$, where $\omega_\bot$ and $\omega_z$ are HO frequencies. In the  examples shown, we focus on prolate-deformed  nuclei ($\eta>1$). We only show results for the  $\sigma=1$ spin component, because of the assumed  time-reversal symmetry. Since the  Coulomb force is neglected, the results for protons and neutrons are identical.

\begin{figure}[htb]
\centering
\includegraphics[width=.8\textwidth]{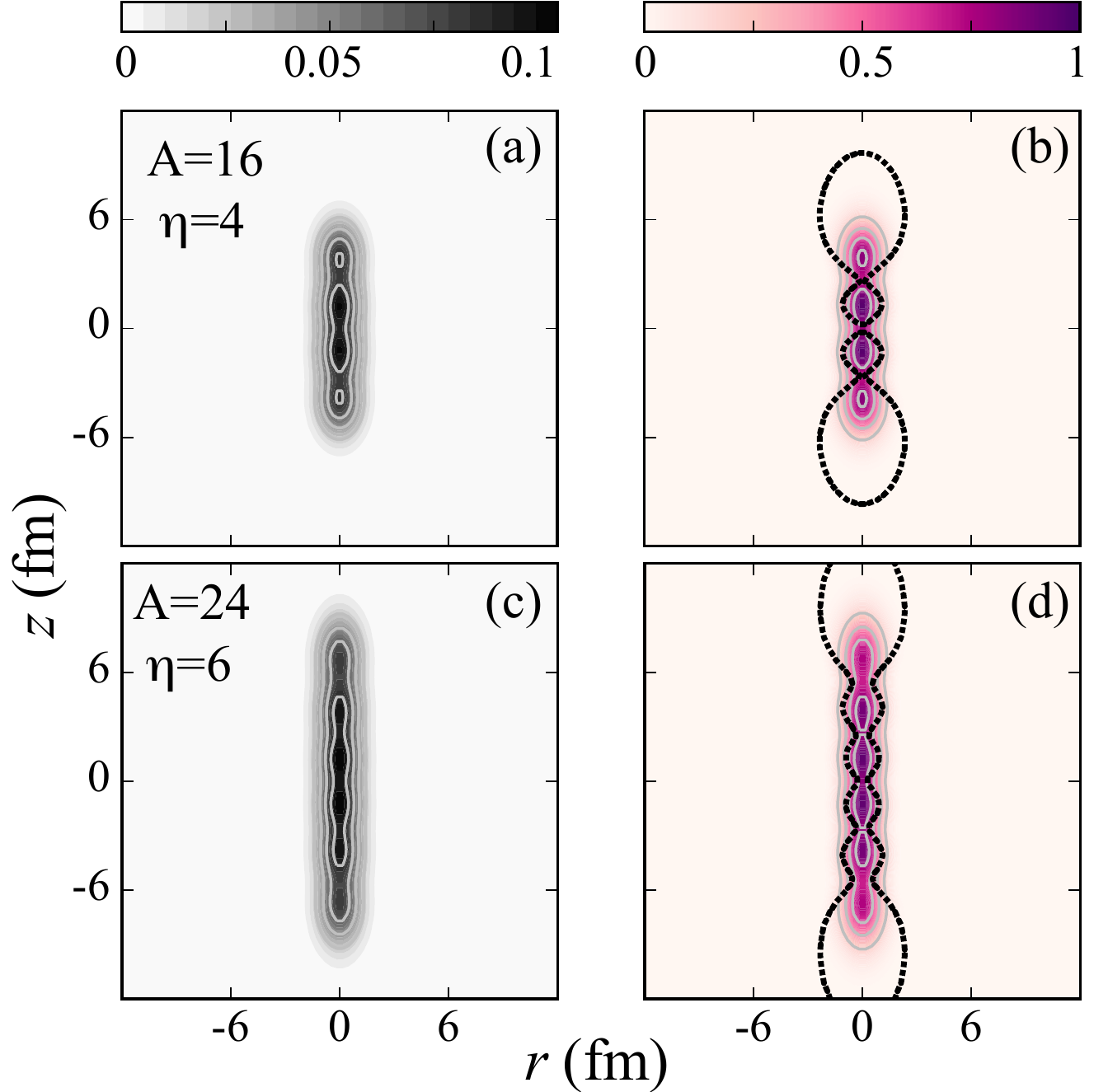} 
\caption{
Linear alpha-chains in the deformed HO model: 4-$\alpha$ chain in  $^{16}$O (top) and  6-$\alpha$ chain in $^{24}$Mg (bottom). Left panels show the particle density of a fixed  spin and isospin (in nucleons/fm$^3$). Right panels show the corresponding NLF (dimensionless), masked with the density form factor  as described in Sec.~\ref{sec:loc}. Black dotted lines are the contour lines of the original definition (\ref{eq:loc}) corresponding to  the value of $\mathcal{C}=0.9$. \label{fig:alpha_chains}}
\end{figure}
In Fig.~\ref{fig:alpha_chains}, we show two typical examples of elongated configurations  in $^{16}$O and $^{24}$Mg. In order to simulate $\alpha$-chain configurations, we choose $\eta$ according to the $\alpha$ particle content. Such a deformation produces the supershell structure associated with the $\eta$-fold SU(3) dynamical symmetry of the rational HO, which stabilizes
the formation of $\alpha$-cluster chains~\cite{Nazarewicz1992}. While a separation into $\alpha$ particles is difficult  to see in the particle density plot, especially for $^{24}$Mg, the NLF clearly reveals four  maxima for $^{16}$O and six  maxima for $^{24}$Mg, with localizations close to one. This means that the nucleons are very localized for each spin/isospin component, implying the presence of aligned $\alpha$ particles. 

\begin{figure}[htb]
\centering
\includegraphics[width=.8\textwidth]{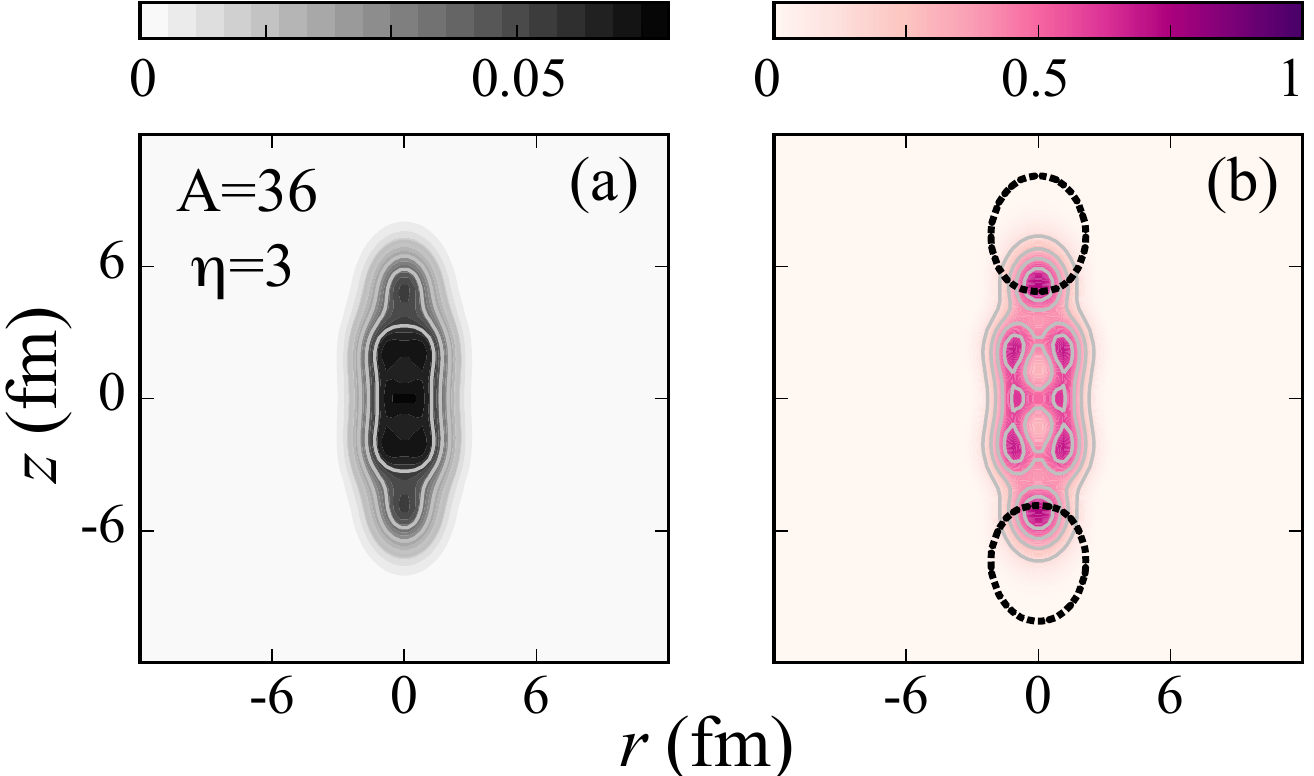}
\caption{Similar to Fig.~\ref{fig:alpha_chains} but for $^{36}$Ar with $\eta=3$.}
\label{fig:ar36-HO}
\end{figure}

As a second example, in Fig.~\ref{fig:ar36-HO} we show a hyperdeformed ($\eta=3$) configuration in $^{36}$Ar. While the particle density hardly shows clustering, the localization shows large values, especially at the tips of the nucleus. The structure in between, corresponds to a deformed $^{28}$Si. It also exhibits cluster structures  at $z=0$ and $z\approx\pm2\fm$. The black dotted line represents the $\mathcal{C}=0.9$ contour  of the original localization measure.

\subsection{DFT description}

\begin{figure}[htb]
\centering
\includegraphics[width=.7\textwidth]{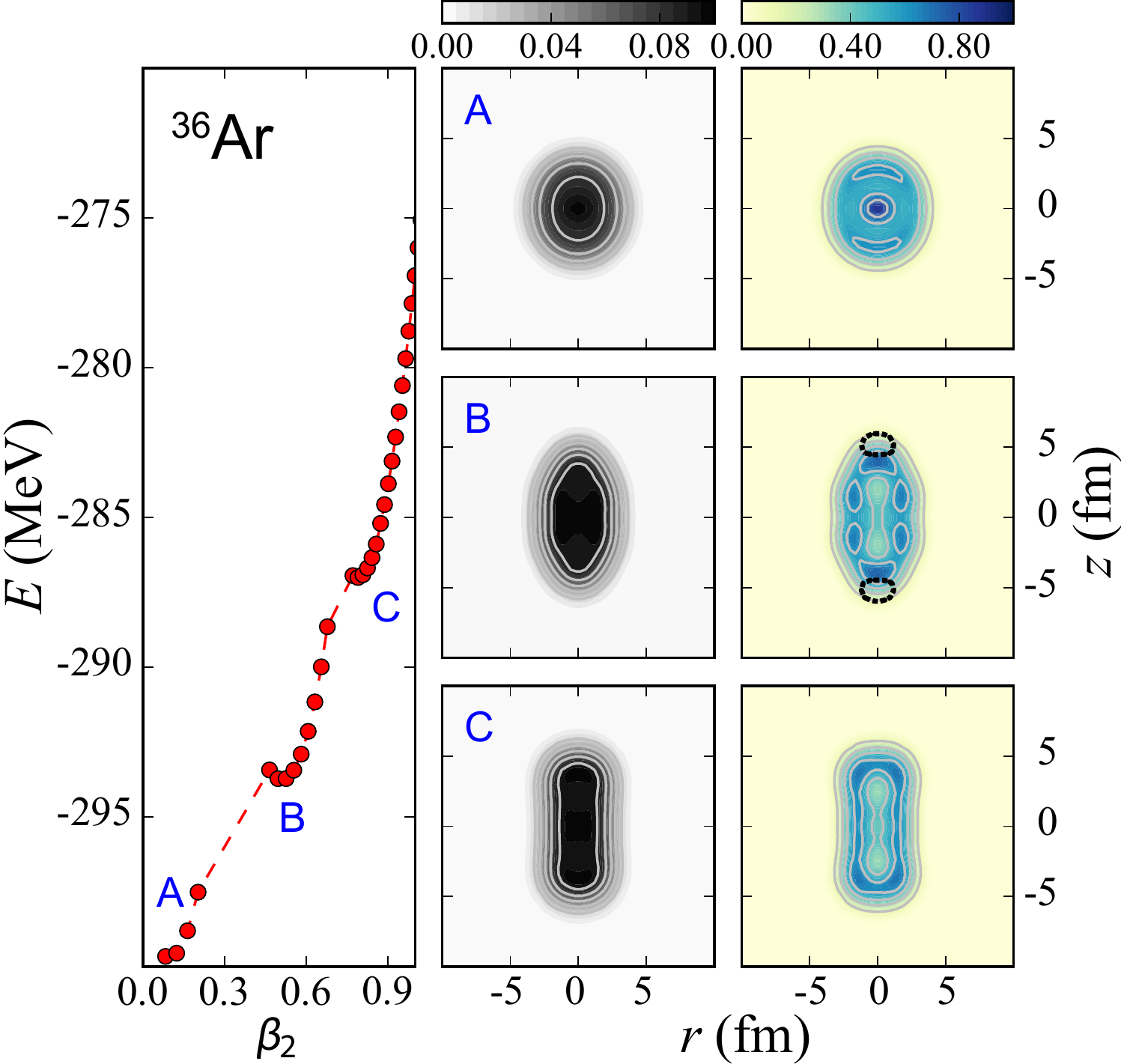}
\caption{Left: Total HF energy of $^{36}$Ar calculated with the functional UNEDF1-HFB versus the quadrupole deformation $\beta_2$. The neutron densities corresponding to the three local minima, marked  A, B, and C are shown in the middle panel ((in nucleons/fm$^3$) and the associated neutron localization is shown in the right panel. Black dotted lines are the contour lines of the original definition (\ref{eq:loc}) corresponding to  the value of $\mathcal{C}=0.9$.}
\label{fig:ar36}
\end{figure}
As a next step we want to compare  the simple HO model predictions with the DFT results. To this end, in Fig.~\ref{fig:ar36} we show the HF energy of $^{36}$Ar as a function of the quadrupole deformation $\beta_2$. The three local minima are predicted at  $\beta_2\approx0.1$, $0.5$,  and $0.8$. The corresponding neutron densities and NLFs are also displayed in  Fig.~\ref{fig:ar36}. (Clustering in $^{36}$Ar has also been studied in the DFT calculations of Ref.~\cite{Ebran2014}.)
The weakly-deformed ground state at $\beta_2\approx0.1$ does not show any structure in the density. Its NLF exhibits a  maximum in the center and an enhancement at the tips. This distribution  constitutes a unique fingerprint of the shell structure of $^{36}$Ar  that is clearly missing in the density plot. The configuration  B is less deformed than that calculated with the HO in Fig.~\ref{fig:ar36-HO}. However, its NLF is similar. In particular, the localization enhancement at the tips reveals  the presence of alpha clustering. The central structure shows  two rings of the enhanced surface localization. Unlike shape B,  shape C has a more uniform NLF, with the localization peaked around the nuclear surface where the contribution from one specific single-particle orbit is likely to dominate.

\begin{figure}[htb]
\centering
\includegraphics[width=.7\textwidth]{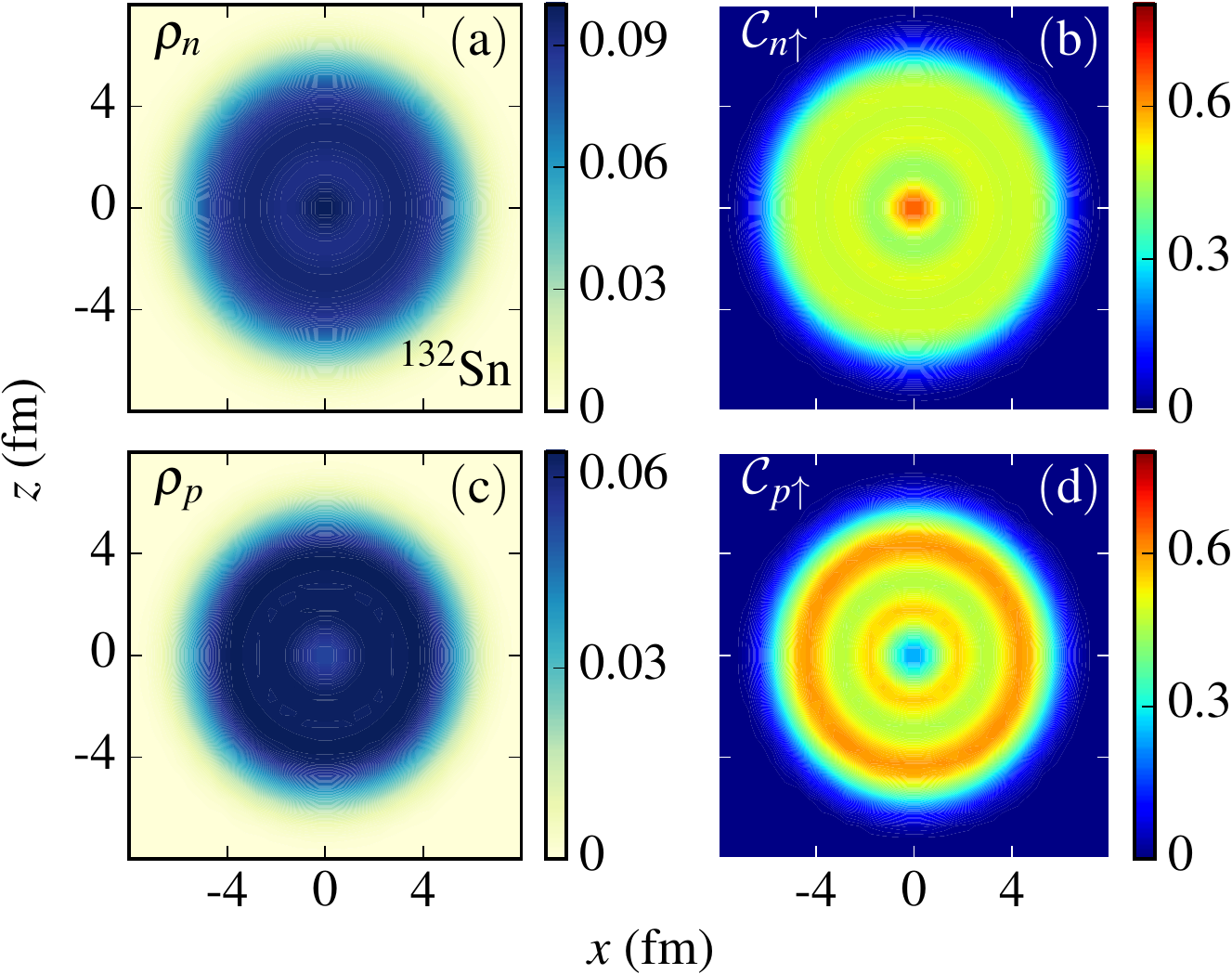}
\caption{HFB calculations for $^{132}$Sn with SkM*: (a) neutron density, (b) neutron localization, (c) proton density, and (d) proton localization. The particle densities are in nucleons/fm$^3$.}
\label{fig:Sn132}
\end{figure}
We now proceed to heavier nuclei. As an illustrative example, we show   in Fig. \ref{fig:Sn132} the distributions computed  for the doubly-magic system  $^{132}$Sn. The particle densities do not exhibit any pronounced shell structure, except perhaps for a small depression of $\rho_p$ in the interior.
The NLFs, however, show  clear patterns of concentric rings of enhanced localization that are distinct for protons and neutrons. While the proton localization has two radial maxima and a central depression, the neutron localization has two radial maxima and  a central maximum. Therefore, the additional closed shell in the neutrons produces an extra region of high localization. Of course, unlike in the atomic systems \cite{Becke}, the total number of shells can not be directly read from the number of peaks in the NLF, because the radial distributions of  wave functions belonging to different nucleonic shells vary fairly smoothly. Nevertheless, as one can assess from Fig. \ref{fig:Sn132}, each magic number leaves a strong imprint on the localization. 

\begin{figure}[htb]
\centering
\includegraphics[width=\textwidth]{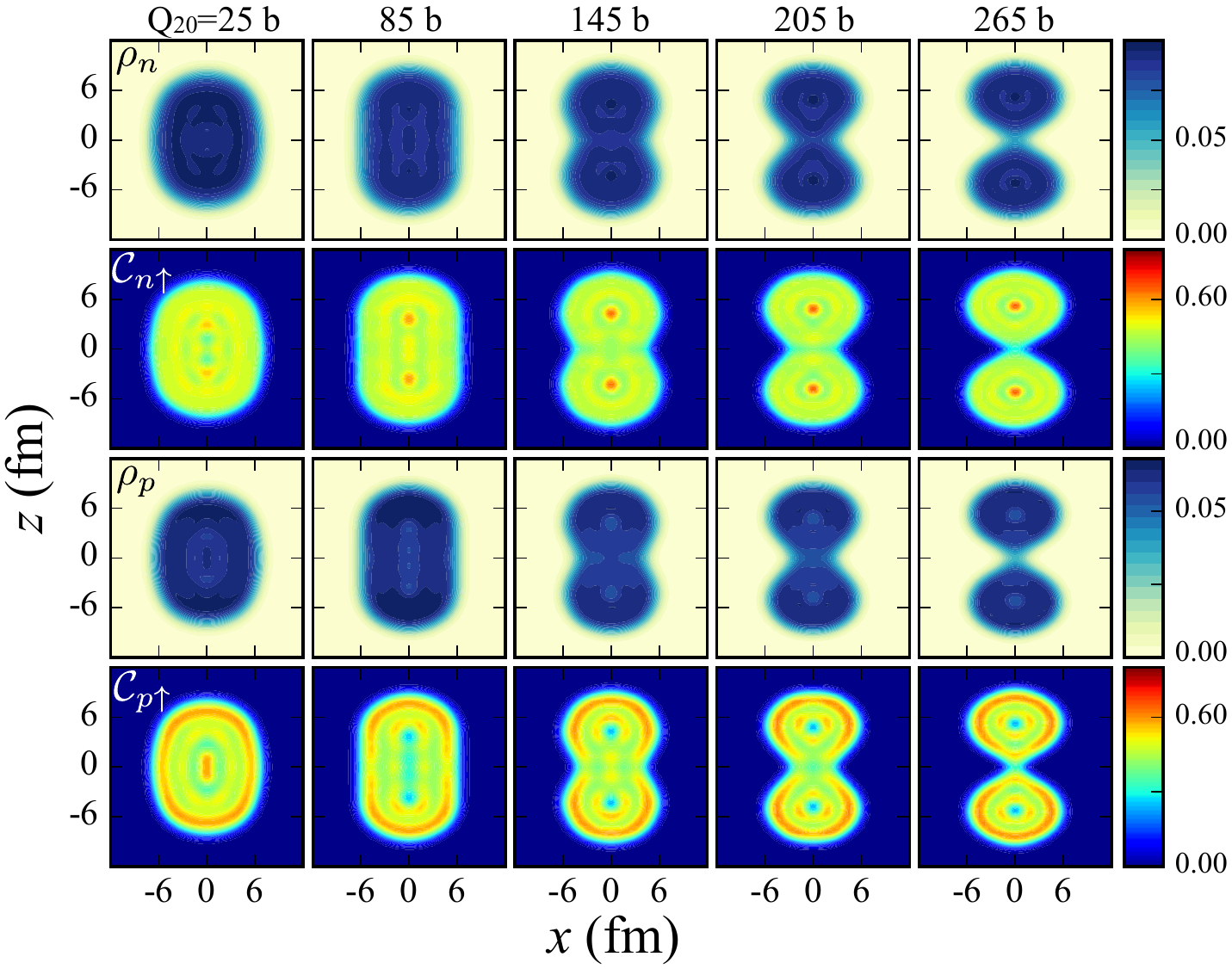}
\caption{HFB calculations for $^{264}$Fm with SkM* for five deformed configurations along the symmetric fission pathway. Shown are: neutron density $\rho_n$ (top); neutron localization $\mathcal{C}_{n\uparrow}$  (second row); proton density $\rho_p$ (third row); and  proton localization $\mathcal{C}_{p\uparrow}$ (bottom). Quadrupole moments $Q_{20}$ characterizing HFB solutions are indicated.   The particle densities are in nucleons/fm$^3$.}
\label{fig:Fm264}
\end{figure}
An application where shell-structure imprints on the NLF are important as they can reveal valuable structural information is nuclear fission. The idea is to recognize the NLF imprint of the fission fragment as it is formed along the fission path. To illustrate this concept, we choose the symmetric fission path of $^{264}$Fm, a subject of several recent DFT studies 
\cite{Staszczak09,Sadhukhan14,Simenel14}.
The results of our HFB SkM* calculations are shown in Fig.~\ref{fig:Fm264}.
As the constraining quadrupole moment $Q_{20}$ gets larger,
the particle densities become increasingly elongated. A neck develops at $Q_{20}\approx 145$\,b, and the scission point is reached at  $Q_{20} \approx 265$\,b above which $^{264}$Fm splits into two  $^{132}$Sn fragments. By comparing the results of Fig.~\ref{fig:Sn132} for $^{132}$Sn one can see the gradual development of the  $^{132}$Sn clusters within the fissioning nucleus.
This example shows in a dramatic way that the NLF can serve as an excellent fingerprint of developing cluster structures in heavy nuclei.

\section{Nuclear rods\label{sec:nucrods}}

\begin{figure}[htb]
\centering
\includegraphics[width=.7\textwidth]{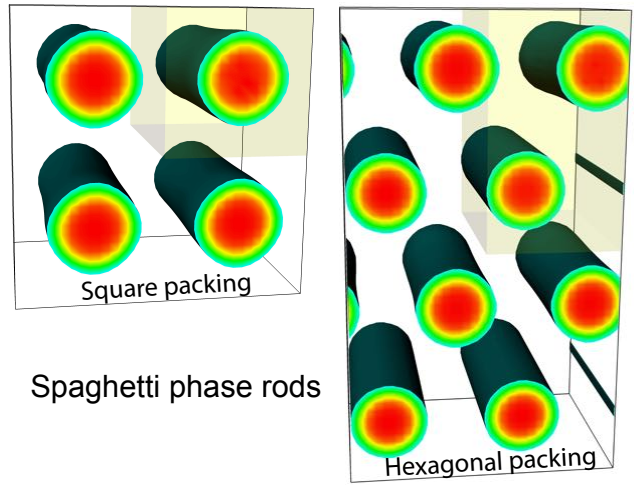} 
\caption{Illustration of square- and hexagonal-packed rod structures. The computational boxes $(L_x,L_y,L_z)$ used are highlighted in yellow. Their dimensions are: (24, 24, 20)\,fm (square packing) and (24, 41.6, 20)\,fm (hexagonal packing).
}
\label{fig:rod-packing}
\end{figure}

In the neutron star crust, or at higher temperatures and higher proton fractions in supernova explosions, exotic pasta phases of nucleonic matter develop. In this study, we use the DFT+TABC method to analyze 
the  nuclear rod (spaghetti) phase as a representative example.  We compare two different packings of rods at a constant mean density $\bar{\rho}=0.0358\fmd$: square and hexagonal packing. Since our computations are performed in rectangular boxes, the unit cell contains one rod for square packing but two rods for hexagonal packing. The computational cells with one periodic repetition of the rods in the $x$- and $y$-directions are shown in Fig.~\ref{fig:rod-packing}. The box length $L_z$ in the $z$-direction, in which the density of the rods is translationally invariant, is sufficiently large so that the finite-volume effects are negligible with TABC. (For a detailed discussion of the performance of TABC for pasta matter, please see \cite{Schuetrumpf2015b}.) The calculations are performed with four different twists in the $z$-direction. In the calculations with no neutron background gas present, a one-dimensional averaging over the twists in the $z$-direction is sufficient. Otherwise a three-dimensional averaging must be used.

Since the geometries are different for square and hexagonal packing, and different numbers of rods are present in a computational box, to compare 
the results of these two variants of calculations we introduce a one-dimensional rod density 
\begin{equation}
\rho_{\rm rod}=\frac{A}{\mbox{\# of rods}\cdot L_z}\quad , \label{eq:rhorod}
\end{equation}
where $A$ is the number of nucleons in the computational box. 
In addition to packing geometry, we also compare predictions of different EDFs:
SLy6 and TOV-min. The parametrization SLy6 \cite{Chabanat} was optimized  to the equation of state of neutron matter and data on magic nuclei; here, constraints were applied to the saturation point of symmetric matter, its compressibility, and its symmetry energy. The EDF TOV-min \cite{Erler2013} was fitted to binding energies, diffraction radii, charge radii and other observables in finite nuclei. Additionally the constraints were imposed on the mass-radius relation for neutron stars to conform with the latest observations; hence, TOV-min is expected to perform well for neutron-rich systems.

\begin{figure}[htb]
\centering
\includegraphics[width=.8\textwidth]{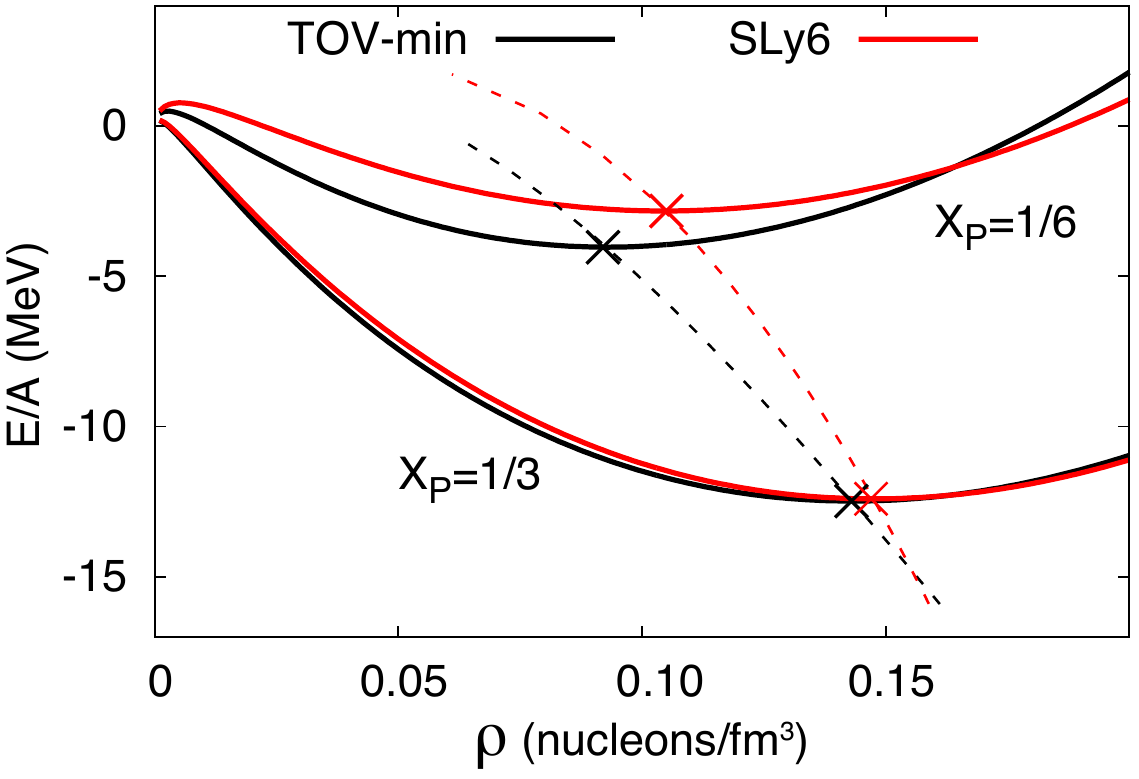}
\caption{Binding energy per nucleon for TOV-min and SLy6 versus the mean density of uniform nuclear matter. The saturation point is marked. The trajectory of saturation points is shown as a dashed line for $X_P\in[0.1,0.5]$}
\label{fig:EoS}
\end{figure}
Figure \ref{fig:EoS} shows the nuclear matter equation of state for TOV-min and SLy6 at the two proton fractions $X_P=1/3$ and $X_P=1/6$, which we use for our rod calculations. While the binding energy is larger for TOV-min at lower proton fractions, both forces are fairly similar at higher values of $X_P$.

\begin{figure}[htp]
\centering
\includegraphics[width=\textwidth]{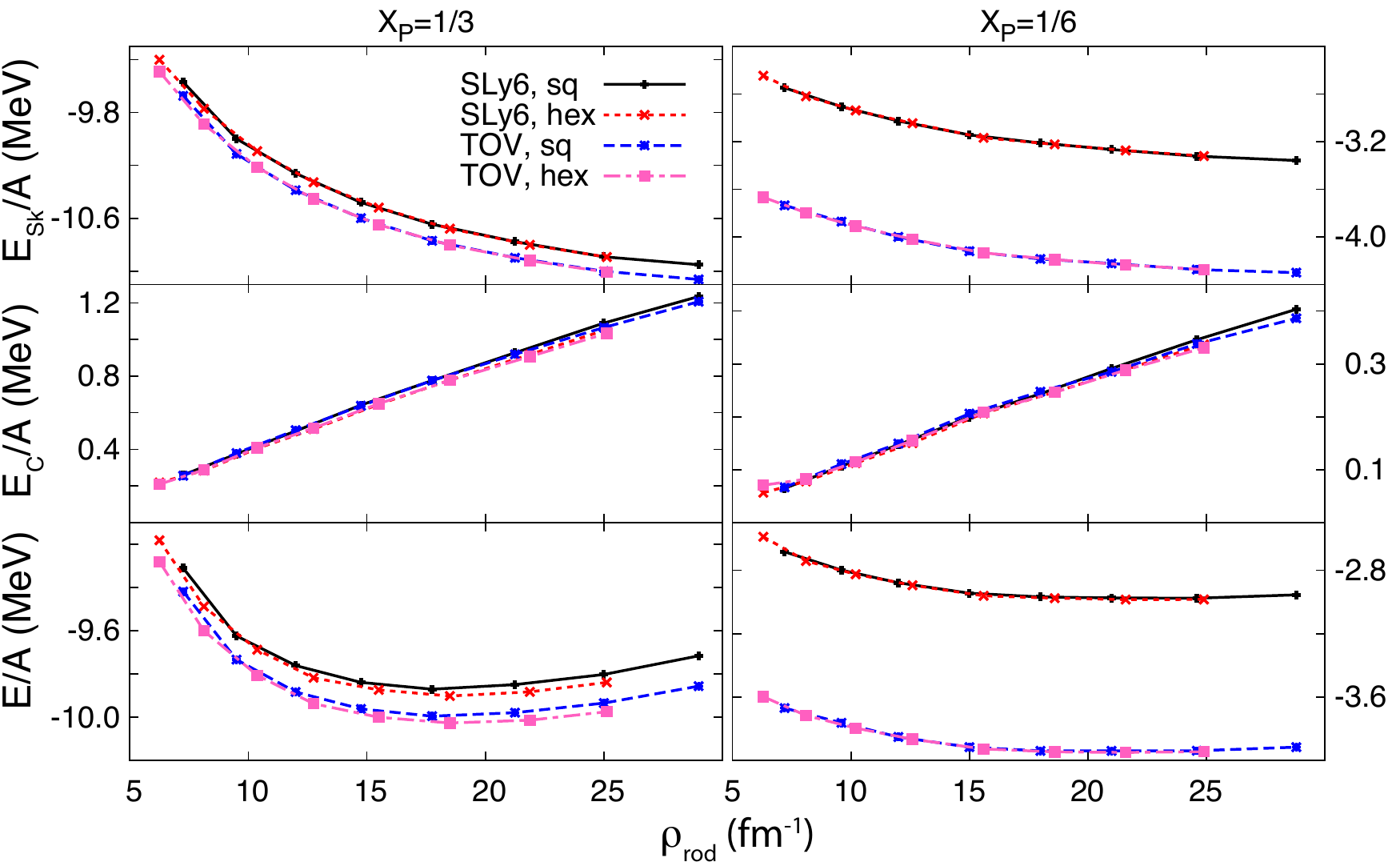}
\caption{Total, Coulomb, and Skyrme energies for squared and hexagonal packed rods in dependence on the 1-D rod density $\rho_{\rm rod}$ as defined in Eq.~\ref{eq:rhorod}. Left panel shows results for $X_P=1/3$ right panel shows $X_P=1/6$. \label{fig:energies} }
\end{figure}

\begin{figure}[htp]
\includegraphics[width=.9\textwidth]{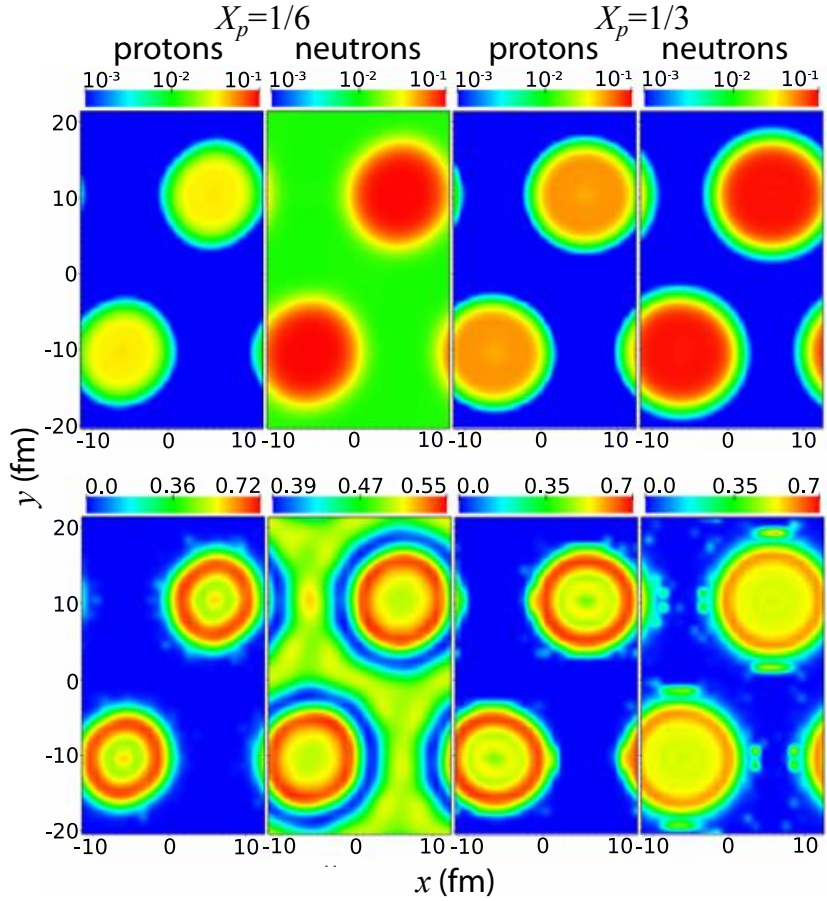}
\caption{Proton and neutron densities (top) and localizations (bottom) at the hexagonal geometry for 
$X_P=1/6$ (left) and $X_P=1/3$ (right)  calculated with DFT+TABC using TOV-min.}
\label{fig:hdens}
\end{figure}

In Fig.~\ref{fig:energies}, the total, Coulomb, and Skyrme energies of SLy6 and TOV-min are shown as a function of $\rho_{\rm rod}$ for the two different values of $X_P$, and different packing geometries. For $X_P=1/3$ both functionals yield similar results, with TOV-min  producing a slightly stronger binding than SLy6.  For the Skyrme EDF, different packings of the rods give very close results, since the rods are not connected. The Skyrme energy is decreasing with larger rods, because the volume term becomes dominant over the surface term. However, the long-ranged Coulomb force is affected by the packing geometry, and the difference is  about 5\% of the total Coulomb energy. The total energy has a minimum near $\rho_{\rm rod}=18\fmk$. 
At the lower proton fraction of $X_P=1/6$ the Skyrme energies strongly depend on the EDF used. The rods calculated with TOV-min are about $1\MeV$ lower in energy than those with SLy6; this is due to the difference in the density dependence of the symmetry energies between these EDFs. At this proton fraction, a neutron background forms, as studied in detail in \cite{Schuetrumpf2014} and is shown in Fig.~\ref{fig:hdens}. The background density is approximately 0.01\,nucleons/fm$^3$. Although the rods are embedded in the neutron gas for both geometries, the energies associated with the nuclear forces, $E_{\rm Sk}$, are very close. This suggests that the neutron background is mostly decoupled from the bound rod structures. As the  total contribution of the Coulomb term is reduced, the impact of different packing geometries is even less important. The minimum of the total energy shifts to the larger rod density, $\rho_{\rm rod}\approx20\fmk$.

Particle densities computed with TOV-min for hexagonal packing are shown in Fig.~\ref{fig:hdens} (top). The results for  SLy6, and for square packing, are very similar; hence, they are not shown.
A small deformation of the proton densities at $X_P=1/6$ can be seen. Most likely, it is a residual finite-volume effect, which has not been fully removed by TABC. For $X_P=1/3$ the twist-averaged shapes are almost perfectly circular. 

The corresponding NLFs are shown in Fig.~\ref{fig:hdens} (bottom). Some irregularities resulting from 
residual finite-volume and grid-discretization effects are seen, as  the localization is a more sensitive measure than the local density.
The localizations exhibit pronounced radial oscillations resulting in ring-like structures, which are fingerprints of shell effects associated with the rod phase.
It is interesting to note that
the neutrons at $X_P=1/6$ show a markedly different localization. Since the neutron gas is present in this case, the localization tends to be around the Fermi-gas value of  ${\cal C}=0.5$. At the inner boundaries of the rods, the localization is enhanced to a value of about 0.55, but right outside the rod, the localization drops to about 0.39. 


\section{Conclusions}\label{Conclusions}

In this work we presented some recent developments pertaining to the DFT description of nucleonic clustering in both finite nuclei and pasta phases. We demonstrated that the nucleonic localization is a superb indicator of clustering in light nuclei, in heavy fissioning systems, and in crustal structures of neutron stars. In particular, the pattern of NLFs can serve as a fingerprint of the single-particle shell structure associated with cluster configurations.

To learn more about pasta structures, we employed the DFT+TABC approach to compare hexagonal and square packing geometries of the rod phase. We conclude that  hexagonal packing is slightly favored by the Coulomb force. The localization of rods depends little on details when no neutron gas is present. In the presence of neutron background, the NLF values become close to the Fermi-gas limit of 0.5, and the structure is washed out, with the main NLF variations happening at the rod's boundaries.

Future studies will include $\alpha$ clustering in medium-mass nuclei as well as  clustering in heavy nuclei, in particular in fission.  Clustering effects and  NLFs will also be studied for various nuclear pasta phases.

\section*{Acknowledgements}

Useful discussions with P.-G. Reinhard and A.S. Umar and the help of E. Olsen are gratefully acknowledged. This material is based upon work supported by the U.S. Department of Energy, Office of Science under Award Numbers DOE-DE-NA0002847 (the Stewardship Science Academic Alliances program), DE-SC0013365 (Michigan State University), and DE-SC0008511 (NUCLEI SciDAC-3 collaboration). This work used computational resources of the Institute for Cyber-Enabled Research at Michigan State University and the National Institute for Computational Sciences (NICS) .


\begin{thebibliography}{59}
\providecommand{\natexlab}[1]{#1}
\providecommand{\url}[1]{\texttt{#1}}
\expandafter\ifx\csname urlstyle\endcsname\relax
  \providecommand{\doi}[1]{doi: #1}\else
  \providecommand{\doi}{doi: \begingroup \urlstyle{rm}\Url}\fi

\bibitem{Beck1}
C.~Beck, ed., \emph{Clusters in Nuclei, vol 1}. vol. 818, \emph{Lecture Notes
  in Physics}, Springer, Berlin, Heidelberg  (2010).

\bibitem{Beck2}
C.~Beck, ed., \emph{Clusters in Nuclei, vol 2}. vol. 848, \emph{Lecture Notes
  in Physics}, Springer, Berlin, Heidelberg  (2012).

\bibitem{Beck3}
C.~Beck, ed., \emph{Clusters in Nuclei, vol 3}. vol. 875, \emph{Lecture Notes
  in Physics}, Springer, Berlin, Heidelberg  (2014).

\bibitem{Oertzen06}
W.~von Oertzen, M.~Freer, and Y.~Kanada-En’yo, Nuclear clusters and nuclear
  molecules, \emph{Phys. Rep.} {\bf 432}, \penalty0 43  (2006).
\newblock \doi{http://dx.doi.org/10.1016/j.physrep.2006.07.001}.

\bibitem{Cluster12}
R.~G. Lovas, Z.~Dombr{\'a}di, G.~G. Kiss, A.~T. Kruppa, and G.~L{\'e}vai, eds.,
  \emph{10th International Conference on Clustering Aspects of Nuclear
  Structure and Dynamics, J. Phys. Conf. Ser.}, vol. 436  (2013).

\bibitem{Ikeda68}
K.~Ikeda, N.~Takigawa, and H.~Horiuchi, The systematic structure-change into
  the molecule-like structures in the self-conjugate 4n nuclei, \emph{Prog.
  Theor. Phys. Suppl.} {\bf E68}, \penalty0 464  (1968).
\newblock \doi{10.1143/PTPS.E68.464}.

\bibitem{Okolowicz12}
J.~Oko{\l}owicz, M.~P{\l}oszajczak, and W.~Nazarewicz, On the origin of nuclear
  clustering, \emph{Prog. Theor. Phys. Suppl.} {\bf 196}, \penalty0 230
  (2012).
\newblock \doi{10.1143/PTPS.196.230}.

\bibitem{Leander}
G.~Leander and S.~Larsson, Potential-energy surfaces for the doubly even {N} = {Z}
  nuclei, \emph{Nucl. Phys. A}. {\bf 239}, \penalty0 93  (1975).
\newblock ISSN 0375-9474.
\newblock \doi{http://dx.doi.org/10.1016/0375-9474(75)91136-7}.

\bibitem{Flocard84}
H.~Flocard, P.~H. Heenen, S.~J. Krieger, and M.~S. Weiss, Configuration space,
  cranked {H}artree-{F}ock calculations for the nuclei 16{O}, 24{M}g and 32{S},
  \emph{Prog. Theor. Phys.} {\bf 72}, \penalty0 1000  (1984).
\newblock \doi{10.1143/PTP.72.1000}.

\bibitem{Freer95}
M.~Freer, R.~Betts, and A.~Wuosmaa, Relationship between the deformed harmonic
  oscillator and clustering in light nuclei, \emph{Nucl. Phys. A}. {\bf 587},
  \penalty0 36  (1995).
\newblock \doi{http://dx.doi.org/10.1016/0375-9474(94)00820-D}.

\bibitem{Ichikawa11}
T.~Ichikawa, J.~A. Maruhn, N.~Itagaki, and S.~Ohkubo, Linear chain structure of
  four-$\ensuremath{\alpha}$ clusters in $^{16}\mathrm{O}$, \emph{Phys. Rev.
  Lett.} {\bf 107}, \penalty0 112501  (2011).
\newblock \doi{10.1103/PhysRevLett.107.112501}.

\bibitem{Ebran12}
J.-P. Ebran, E.~Khan, T.~Nik\v{s}i{\'c}, and D.~Vretenar, How atomic nuclei
  cluster, \emph{Nature}. {\bf 487}, \penalty0 341  (2012).

\bibitem{Ebran2014}
J.-P. Ebran, E.~Khan, T.~Nik\ifmmode \check{s}\else
  \v{s}\fi{}i\ifmmode~\acute{c}\else \'{c}\fi{}, and D.~Vretenar, Density
  functional theory studies of cluster states in nuclei, \emph{Phys. Rev. C}.
  {\bf 90}, \penalty0 054329  (2014).
\newblock \doi{10.1103/PhysRevC.90.054329}.

\bibitem{Hecht77}
K.~T. Hecht, Relation between cluster and shell-model wave functions,
  \emph{Phys. Rev. C}. {\bf 16}, \penalty0 2401  (1977).
\newblock \doi{10.1103/PhysRevC.16.2401}.

\bibitem{Nazarewicz1992}
W.~Nazarewicz and J.~Dobaczewski, Dynamical symmetries, multiclustering, and
  octupole susceptibility in superdeformed and hyperdeformed nuclei,
  \emph{Phys. Rev. Lett.} {\bf 68}, \penalty0 154  (1992).
\newblock \doi{10.1103/PhysRevLett.68.154}.

\bibitem{(Elh15)}
S.~Elhatisari, D.~Lee, G.~Rupak, E.~Epelbaum, H.~Krebs, T.~A. L{\"a}hde,
  T.~Luu, and U.-G. Mei{\ss}ner, Ab initio alpha--alpha scattering,
  \emph{Nature}. {\bf 528}, \penalty0 111  (2015).
\newblock \doi{10.1038/nature16067}.

\bibitem{Bender03}
M.~Bender, P.-H. Heenen, and P.-G. Reinhard, Self-consistent mean-field models
  for nuclear structure, \emph{Rev. Mod. Phys.} {\bf 75}, \penalty0 121--180
  (2003).
\newblock \doi{10.1103/RevModPhys.75.121}.

\bibitem{Becke}
A.~D. Becke and K.~E. Edgecombe, A simple measure of electron localization in
  atomic and molecular systems, \emph{J Chem. Phys.} {\bf 92}, \penalty0 5397
  (1990).

\bibitem{Savin97}
A.~Savin, R.~Nesper, S.~Wengert, and T.~F. F{\"a}ssler, {ELF}: The electron
  localization function, \emph{Angew. Chem. Int. Ed. Engl.} {\bf 36}, \penalty0
  1808  (1997).
\newblock \doi{10.1002/anie.199718081}.

\bibitem{Scemama}
A.~Scemama, P.~Chaquin, and M.~Caffarel, Electron pair localization function: A
  practical tool to visualize electron localization in molecules from quantum
  monte carlo data, \emph{J. Chem. Phys.} {\bf 121}, \penalty0 1725  (2004).
\newblock \doi{http://dx.doi.org/10.1063/1.1765098}.

\bibitem{Burnus05}
T.~Burnus, M.~A.~L. Marques, and E.~K.~U. Gross, Time-dependent electron
  localization function, \emph{Phys. Rev. A}. {\bf 71}, \penalty0 010501  (2005).
\newblock \doi{10.1103/PhysRevA.71.010501}.

\bibitem{Reinhard2011}
P.-G. Reinhard, J.~A. Maruhn, A.~S. Umar, and V.~E. Oberacker, Localization in
  light nuclei, \emph{Phys. Rev. C}. {\bf 83}, \penalty0 034312  (2011).
\newblock \doi{10.1103/PhysRevC.83.034312}.

\bibitem{Ravenhall}
D.~G. Ravenhall, C.~J. Pethick, and J.~R. Wilson, Structure of matter below
  nuclear saturation density, \emph{Phys. Rev. Lett.} {\bf 50}, \penalty0
  2066--2069  (1983).
\newblock \doi{10.1103/PhysRevLett.50.2066}.

\bibitem{Hashimoto}
M.~Hashimoto, H.~Seki, and M.~Yamada, Shape of nuclei in the crust of neutron
  star, \emph{Prog. Theor. Phys.} {\bf 71}\penalty0 (2), \penalty0 320--326
  (1984).
\newblock \doi{10.1143/PTP.71.320}.

\bibitem{Dorso2012}
C.~O. Dorso, P.~A. Gim\'enez~Molinelli, and J.~A. L\'opez, Topological
  characterization of neutron star crusts, \emph{Phys. Rev. C}. {\bf 86},
  \penalty0 055805  (2012).
\newblock \doi{10.1103/PhysRevC.86.055805}.

\bibitem{Williams1985}
R.~Williams and S.~Koonin, Sub-saturation phases of nuclear matter, \emph{Nucl.
  Phys.} {\bf 435}\penalty0 (3?4), \penalty0 844 -- 858  (1985).
\newblock ISSN 0375-9474.
\newblock \doi{10.1016/0375-9474(85)90191-5}.

\bibitem{Oka13a}
M.~Okamoto, T.~Maruyama, K.~Yabana, and T.~Tatsumi, Nuclear ``pasta''
  structures in low-density nuclear matter and properties of the neutron-star
  crust, \emph{Phys. Rev. C}. {\bf 88}, \penalty0 025801  (2013).

\bibitem{Pais15}
H.~Pais, S.~Chiacchiera, and C.~Provid{\^e}ncia, Light clusters, pasta phases,
  and phase transitions in core-collapse supernova matter, \emph{Phys. Rev. C}.
  {\bf 91}, \penalty0 055801  (2015).
\newblock \doi{10.1103/PhysRevC.91.055801}.

\bibitem{Schneider13}
A.~S. Schneider, C.~J. Horowitz, J.~Hughto, and D.~K. Berry, Nuclear ``pasta''
  formation, \emph{Phys. Rev. C}. {\bf 88}, \penalty0 065807  (2013).
\newblock \doi{10.1103/PhysRevC.88.065807}.

\bibitem{Schneider14}
A.~S. Schneider, D.~K. Berry, C.~M. Briggs, M.~E. Caplan, and C.~J. Horowitz,
  Nuclear ``waffles'', \emph{Phys. Rev. C}. {\bf 90}, \penalty0 055805  (2014).

\bibitem{Maruyama}
T.~Maruyama et~al., Quantum molecular dynamics approach to the nuclear matter
  below the saturation density, \emph{Phys. Rev. C}. {\bf 57}, \penalty0
  655--665  (Feb, 1998).
\newblock \doi{10.1103/PhysRevC.57.655}.

\bibitem{Sonoda2008}
H.~Sonoda, G.~Watanabe, K.~Sato, K.~Yasuoka, and T.~Ebisuzaki, Phase diagram of
  nuclear ``pasta'' and its uncertainties in supernova cores, \emph{Phys. Rev.
  C}. {\bf 77}, \penalty0 035806  (Mar, 2008).
\newblock \doi{10.1103/PhysRevC.77.035806}.

\bibitem{Watanabe09}
G.~Watanabe, H.~Sonoda, T.~Maruyama, K.~Sato, K.~Yasuoka, and T.~Ebisuzaki,
  Formation of nuclear ``pasta'' in supernovae, \emph{Phys. Rev. Lett.} {\bf
  103}, \penalty0 121101  (2009).

\bibitem{Bonche}
P.~Bonche and D.~Vautherin, A mean-field calculation of the equation of state
  of supernova matter, \emph{Nucl. Phys. A}. {\bf 372}\penalty0 (3), \penalty0
  496 -- 526  (1981).
\newblock ISSN 0375-9474.
\newblock \doi{10.1016/0375-9474(81)90049-X}.

\bibitem{Mag02}
P.~Magierski and P.-H. Heenen, Structure of the inner crust of neutron stars:
  Crystal lattice or disordered phase?, \emph{Phys. Rev. C}. {\bf 65},
  \penalty0 045804  (2002).
\newblock \doi{10.1103/PhysRevC.65.045804}.

\bibitem{Goegelein}
P.~G\"ogelein and H.~M\"uther, Nuclear matter in the crust of neutron stars,
  \emph{Phys. Rev. C}. {\bf 76}, \penalty0 024312  (2007).
\newblock \doi{10.1103/PhysRevC.76.024312}.

\bibitem{NewtonStone}
W.~G. Newton and J.~R. Stone, Modeling nuclear ``pasta'' and the transition to
  uniform nuclear matter with the 3d {S}kyrme-{H}artree-{F}ock method at finite
  temperature: Core-collapse supernovae, \emph{Phys. Rev. C}. {\bf 79},
  \penalty0 055801  (2009).

\bibitem{Pais12}
H.~Pais and J.~R. Stone, Exploring the nuclear pasta phase in core-collapse
  supernova matter, \emph{Phys. Rev. Lett.} {\bf 109}, \penalty0 151101  (2012).
\newblock \doi{10.1103/PhysRevLett.109.151101}.

\bibitem{Schuetrumpf2013a}
B.~Schuetrumpf, M.~A. Klatt, K.~Iida, J.~A. Maruhn, K.~Mecke, and P.-G.
  Reinhard, Time-dependent {H}artree-{F}ock approach to nuclear ``pasta'' at finite
  temperature, \emph{Phys. Rev. C}. {\bf 87}, \penalty0 055805  (2013).
\newblock \doi{10.1103/PhysRevC.87.055805}.

\bibitem{Schuetrumpf2014}
B.~Schuetrumpf, K.~Iida, J.~A. Maruhn, and P.-G. Reinhard, Nuclear ``pasta
  matter'' for different proton fractions, \emph{Phys. Rev. C}. {\bf 90},
  \penalty0 055802  (2014).
\newblock \doi{10.1103/PhysRevC.90.055802}.

\bibitem{Schuetrumpf2015}
B.~Schuetrumpf, M.~A. Klatt, K.~Iida, G.~E. Schr\"oder-Turk, J.~A. Maruhn,
  K.~Mecke, and P.-G. Reinhard, Appearance of the single gyroid network phase
  in ``nuclear pasta'' matter, \emph{Phys. Rev. C}. {\bf 91}, \penalty0 025801
  (2015).
\newblock \doi{10.1103/PhysRevC.91.025801}.

\bibitem{Gros1992}
C.~Gros, The boundary condition integration technique: results for the hubbard
  model in 1d and 2d, \emph{Z. Phys. B}. {\bf 86}\penalty0 (3), \penalty0
  359--365  (1992).

\bibitem{Gros1996}
C.~Gros, Control of the finite-size corrections in exact diagonalization
  studies, \emph{Phys. Rev. B}. {\bf 53}, \penalty0 6865--6868  (1996).
\newblock \doi{10.1103/PhysRevB.53.6865}.

\bibitem{Lin}
C.~Lin, F.~H. Zong, and D.~M. Ceperley, Twist-averaged boundary conditions in
  continuum quantum monte carlo algorithms, \emph{Phys. Rev. E}. {\bf 64},
  \penalty0 016702  (2001).
\newblock \doi{10.1103/PhysRevE.64.016702}.

\bibitem{Carter}
B.~Carter, N.~Chamel, and P.~Haensel, Entrainment coefficient and effective
  mass for conduction neutrons in neutron star crust: simple microscopic
  models, \emph{Nuclear Physics A}. {\bf 748}\penalty0 (3--4), \penalty0 675 --
  697  (2005).
\newblock ISSN 0375-9474.
\newblock \doi{http://dx.doi.org/10.1016/j.nuclphysa.2004.11.006}.

\bibitem{Chamel2007}
N.~Chamel, S.~Naimi, E.~Khan, and J.~Margueron, Validity of the {W}igner-{S}eitz
  approximation in neutron star crust, \emph{Phys. Rev. C}. {\bf 75}, \penalty0
  055806  (2007).
\newblock \doi{10.1103/PhysRevC.75.055806}.

\bibitem{Schuetrumpf2015b}
B.~Schuetrumpf and W.~Nazarewicz, Twist-averaged boundary conditions for
  nuclear pasta {H}artree-{F}ock calculations, \emph{Phys. Rev. C}. {\bf 92},
  \penalty0 045806  (2015).
\newblock \doi{10.1103/PhysRevC.92.045806}.

\bibitem{Gulminelli}
F.~Gulminelli, T.~Furuta, O.~Juillet, and C.~Leclercq, Boundary conditions for
  star matter and other periodic fermionic systems, \emph{Phys. Rev. C}. {\bf
  84}, \penalty0 065806  (2011).
\newblock \doi{10.1103/PhysRevC.84.065806}.

\bibitem{SkMS}
J.~Bartel, P~ Quentin, M.~Brack, C.~Guet, and H.-B.~H\r{a}kansson, Towards a better 
parametrisation of Skyrme-like effective forces: A critical study of the SkM force, 
\emph{Nucl. Phys. A}. {\bf 386}, \penalty0 (1), \penalty0 79 -- 100  (1982).
\newblock ISSN 0375-9474

\bibitem{UNEDF1HFB}
N.~Schunck, J.~D. McDonnell, J.~Sarich, S.~M. Wild, and D.~Higdon, Error
  analysis in nuclear density functional theory, \emph{J. Phys. G}. {\bf
  42}\penalty0 (3), \penalty0 034024  (2015).

\bibitem{Chabanat}
E.~Chabanat et~al., A skyrme parametrization from subnuclear to neutron star
  densities part ii. nuclei far from stabilities, \emph{Nucl. Phys. A}. {\bf
  635}\penalty0 (1--2), \penalty0 231 -- 256  (1998).
\newblock ISSN 0375-9474.

\bibitem{Erler2013}
J.~Erler, C.~J. Horowitz, W.~Nazarewicz, M.~Rafalski, and P.-G. Reinhard,
  Energy density functional for nuclei and neutron stars, \emph{Phys. Rev. C}.
  {\bf 87}, \penalty0 044320  (2013).
\newblock \doi{10.1103/PhysRevC.87.044320}.

\bibitem{Stoitsov2013}
M.~Stoitsov, N.~Schunck, M.~Kortelainen, N.~Michel, H.~Nam, E.~Olsen,
  J.~Sarich, and S.~Wild, Axially deformed solution of the
  {S}kyrme-{H}artree--{F}ock--{B}ogoliubov equations using the transformed harmonic
  oscillator basis ({II}) hfbtho v2.00d: A new version of the program,
  \emph{Computer Physics Communications}. {\bf 184}\penalty0 (6), \penalty0
  1592 -- 1604  (2013).
\newblock ISSN 0010-4655.
\newblock \doi{http://dx.doi.org/10.1016/j.cpc.2013.01.013}.

\bibitem{Schunck2012}
N.~Schunck, J.~Dobaczewski, J.~McDonnell, W.~Satu{\l}a, J.~Sheikh,
  A.~Staszczak, M.~Stoitsov, and P.~Toivanen, Solution of the
  {S}kyrme--{H}artree--{F}ock--{B}ogolyubov equations in the cartesian deformed
  harmonic-oscillator basis.: ({VII}) {HFODD} ({V}2.49t): A new version of the
  program, \emph{Computer Physics Communications}. {\bf 183}\penalty0 (1),
  \penalty0 166 -- 192  (2012).
\newblock ISSN 0010-4655.
\newblock \doi{http://dx.doi.org/10.1016/j.cpc.2011.08.013}.

\bibitem{Mar15a}
J.~Maruhn, P.-G. Reinhard, P.~Stevenson, and A.~Umar, The {TDHF} code sky3d,
  \emph{Comput. Phys. Commun.} {\bf 185}\penalty0 (7), \penalty0 2195 -- 2216
  (2014).
\newblock ISSN 0010-4655.
\newblock \doi{http://dx.doi.org/10.1016/j.cpc.2014.04.008}.

\bibitem{Rae88}
W.~Rae, Clustering phenomena and shell effects in nuclear structure and
  reactions, \emph{Int. J. Mod. Phys. A}. {\bf 03}, \penalty0 1343  (1988).
\newblock \doi{10.1142/S0217751X88000576}.

\bibitem{Staszczak09}
A.~Staszczak, A.~Baran, J.~Dobaczewski, and W.~Nazarewicz, Microscopic
  description of complex nuclear decay: Multimodal fission, \emph{Phys. Rev.
  C}. {\bf 80}, \penalty0 014309  (2009).
\newblock \doi{10.1103/PhysRevC.80.014309}.

\bibitem{Sadhukhan14}
J.~Sadhukhan, J.~Dobaczewski, W.~Nazarewicz, J.~A. Sheikh, and A.~Baran,
  Pairing-induced speedup of nuclear spontaneous fission, \emph{Phys. Rev. C}.
  {\bf 90}, \penalty0 061304  (2014).
\newblock \doi{10.1103/PhysRevC.90.061304}.

\bibitem{Simenel14}
C.~Simenel and A.~S. Umar, Formation and dynamics of fission fragments,
  \emph{Phys. Rev. C}. {\bf 89}, \penalty0 031601  (2014).
\newblock \doi{10.1103/PhysRevC.89.031601}.

\end{thebibliography}

\end{document}